\documentstyle[aps,prl,manuscript,epsf,rotate]{revtex}

\begin{document}

\title{
\rightline{\small KANAZAWA 98-18}
Correlated Instanton Orientations 
in the $SU(2)$ Yang-Mills Vacuum 
and Pair Formation in the Deconfined Phase
}
\author{
    Ernst-Michael Ilgenfritz$^{1}$, Stefan Thurner$^{2}$\\
    $^{1}${\it Institute for Theoretical Physics, Kanzawa University, Japan} \\
    $^{2}${\it Institut f\"ur Kernphysik, TU-Wien, Austria }\\
}

\maketitle

\begin{abstract}
Using the renormalization group motivated smoothing technique
we study the semiclassical structure of the pure Yang-Mills
vacuum. We carefully check that identified clusters of topological
charge behave like instantons around their centers.
Looking at distance distributions of clusters we find
instanton-antiinstanton pair formation in the deconfined phase.
We use suitably normalized gauge invariant field strength correlators
to obtain first quantitative information on the relative
color orientation of instantons.
We give further evidence for a modified instanton profile
in the confined phase and present a simple alternative model 
for it. 

\vfill

\noindent
PACS number(s):
11.15.Kc,  
12.38.Gc,  
11.10.Wx,  
11.10.Hi  


\end{abstract}

\newpage 

The last decade has witnessed tremendous success in the practical
application of the instanton model to describe 
hadronic physics~\cite{SHURYAK93,DIAKONOV96,SCHAEFER98}.
This model is a radical simplification of QCD, taking just one
specific feature of Yang-Mills theory into account, the existence of
instantons~\cite{BPST}. Within this framework it is very natural 
to understand chiral symmetry breaking~\cite{CALLAN78,NONQUENCHED}
(and its restoration at the hadronic phase transition) and to  
solve the $U_A(1)$ problem~\cite{tHOOFT'U1}, 
but confinement has defied an explanation so far.
The model has been formulated to represent a practical tool 
for a wide range of non-perturbative calculations at zero and non-zero 
temperature. Instanton models have been developed ranging  
from the random (uncorrelated) ideal instanton liquid model 
RILM~\cite{RILM}, used mainly for zero temperature,
to the interacting (correlated) instanton liquid model IILM~\cite{IILM}, 
in particular for $T\ne0$, when the temperature
dependence of the interaction becomes important~\cite{ILGENFRITZ94}. 
Even in the RILM there is some trace of interaction 
in so far as without some repulsive interaction~\cite{ILGENFRITZ81} 
the existence of a distinguished instanton size could not be postulated.

Presently, some cornerstones of the IILM become subject of interest for
lattice investigations and get immediate phenomenological importance. 
In deep inelastic scattering (DIS), attractive 
instanton-antiinstanton (valley) configurations within a narrow
range of relative orientations, and the small size flank of the
size distribution $d(\rho)$ are substantial for the event rate estimated for 
instanton mediated multiparticle processes~\cite{RINGWALD98}. 
Recently there was some effort to provide parts of this 
information, defining instanton size and distance 
distributions on the lattice, an enterprise which is under 
current debate~\cite{FORCRAND97,KOVACS98,NEGELE98}. 

The basic idea of instanton models is to reduce the infinite 
number of integration variables in the path integral,
to an integral over a manageable set of relatively few collective coordinates. 
Those are attributed to instantons  
forming an interacting system that is described by the grand-canonical 
ensemble with the partition function
\begin{equation}
\label{instZ}
 Z = \sum_{N_{+},N_{-}} \frac{1}{N_{+}!N_{-}!}
 \prod_i^{N_{+}}\int [d\Omega_i^{+}\, n(\rho_i^{+})]
 \prod_j^{N_{-}}\int [d\Omega_j^{-}\, n(\rho_j^{-})]
  \, \exp(-S_{\mathrm{int}}).
\end{equation}
Here 
$\Omega_i^{\epsilon}=(z_i^{\epsilon},\rho_i^{\epsilon},U_i^{\epsilon})$ 
are the collective coordinates of
the $N_{+}$ instantons ($\epsilon_i=+$) 
and $N_{-}$ antiinstantons ($\epsilon_i=-$); 
$z_i^{\epsilon}$ are the center positions, 
$\rho_i^{\epsilon}$ the sizes and the $SU(N_c)$ matrices 
$U_i^{\epsilon}$ the color orientations of (anti-)instantons. 
$n(\rho)$ is the semiclassical instanton distribution function~\cite{tHOOFT76} 
\begin{equation}
\label{amplitude}
 dn_I = \frac{0.466\exp(-1.679 N_c)}{(N_c-1)!(N_c-2)!}
  \left(\frac{8 \pi^2}{g^2}\right)^{2 N_c}
   \exp\left(-\frac{8\pi^2}{g^2(\rho)}\right)
    \frac{d^4zd\rho}{\rho^5}.
\end{equation}
The instanton interaction $S_{int}$ can partly be taken into account in
a cut-off single-instanton density function. The form
\begin{equation}
\label{density}
d(\rho) = \frac{1}{\rho^5} (\rho \Lambda)^b
         \exp \left[-\nu \frac{\rho^2}{\overline{\rho^2}}\right],
\end{equation}
results from a mean-field approximation~\cite{DIAKONOV84} due to the 
instanton-instanton and instanton-antiinstanton interaction, being 
repulsive on average. $b$ is the one-loop coefficient of the QCD 
$\beta$-function ($b=11N_c/3$ for pure Yang-Mills theory), and $\nu=(b-4)/2$. 
This results in a packing fraction
$ \frac{N}{V}~\overline{\rho^2}^2=\frac{\nu}{\beta ~\gamma^2}  $, 
where the instanton interaction is measured by 
$\gamma^2 = \frac{27}{4}\frac{N_c}{N_c^2-1}\pi^2$ and 
$\beta=\beta(\overline{\rho})\approx 15$ the average action per instanton.  
The color factor $\gamma^2$ represents the averaging over (relative) 
gauge orientations. 

To compute mesonic correlation functions in this framework, Bethe-Salpeter 
equations~\cite{HUTTER} can be formulated in the so-called random phase 
approximation. For this purpose it is assumed that the instanton orientation is 
random and the distribution of instanton centers is uncorrelated.
Baryonic correlation functions are difficult to obtain in this way.
For crude  estimates of the condensates, the knowledge of  
the instanton size distribution $d(\rho)$ (or simply the average $\rho$) 
and the distribution of distances (or the average nearest neighbor 
distance) is sufficient. Much of the knowledge about hadronic
correlation functions~\cite{SHURYAK93} and about the spectral 
density~\cite{CHIRAL} of the Dirac operator, however,  
has been obtained numerically, from Monte Carlo simulations of the
instanton-antiinstanton liquid. Within this method of
calculation it is possible to incorporate information on pair
distributions in Euclidean and color orientation space which originates 
from specific Ans\"atze or is provided by lattice observations.

Very little was known, so far, about the relative color orientation.
Concerning the quark mediated interactions, the fermionic determinant 
which enters a {\it non-quenched instanton} Monte Carlo~\cite{NONQUENCHED}
can be reasonably parametrized in terms of the zero-mode overlap matrix element 
$T_{IA} \propto i (u_{\mathrm{rel}}\Delta z)\rho^{+}\rho^{-}/
	     (2\rho^{+}\rho^{-}+(\Delta z)^2)^2$ 
in order to sample intanton orientations {\it \'a la} 
Metropolis~\footnote{$u_{\mathrm{rel}}$ is the 
real 4-vector which represents the $SU(2)$ matrix
$U_{\mathrm{rel}}$ and $\Delta z = z^{+}-z^{-}$.}.
The gluonic part of the interaction is even less known, {\it i.e.} strongly 
dependent on the Ansatz of II or IA superposition.
Usually it is taken care of by a sharp $\rho$-distribution (\ref{density}) and 
by a hard-core cut-off~\cite{ILGENFRITZ81}. The most important and remarkably 
Ansatz independent source of orientational interaction is the dipole-dipole 
interaction which acts exclusively between instanton and anti-instanton and 
does not contribute to the coefficient $\gamma$ because the mean-field 
approximation neglects color correlations.

In this letter we give first quantitative results on the color 
correlation in instanton-instanton and instanton-antiinstanton pairs 
in pure Yang-Mills theory, which exists entirely due to gluonic interaction. 
Further we present the pair distribution between topological clusters as a 
function of distance in $4D$ space. Finally, we critically examine the 
assumed color coherence within a topological cluster.
The progress in this paper towards a more detailed description of
the instanton vacuum structure would not be possible without
the method of constrained smoothing (CS) which makes
use of the concept of perfect lattice action~\cite{HASENFRATZ94}. It allows 
to study semiclassical configurations which can be far 
from being exact solutions of the Euclidean field equations. 
Our Monte Carlo configurations are created with the same action
using the Metropolis method.  The CS procedure, in contrast to cooling
or smearing, is essentially a two-lattice procedure, where lattice
configurations are related to each other by blocking (fine-to-coarse) 
and inverse blocking~\cite{HASENFRATZ94} (coarse-to-fine). 
The smoothed configuration 
on the fine lattice, emerging from inverse blocking a blocked configuration,
can be considered as {\it semiclassical}, which embraces deformations from 
classical solutions due to classical and quantum interactions.
In the present work we want to get information about these interactions.
In particular, nearby instanton-antiinstanton (AI) pairs  will not be 
annihilated by CS. The distance and color correlations between these objects 
(as well as between instanton-instanton pairs) is not known from rigorous 
analytical calculations. Another possible deviation that we expect 
is a non-classical behavior of the field strength (lack of coherence)
within a topological cluster which could give rise to a modified profile 
of an instanton.
In contrast to CS, cooling does not converge before {\it locally classical 
fields} have been obtained. Therefore, a critical assessment of instantons 
in the quantum ensemble seems to be impossible. 
After some number of iterations, 
cooling or smearing destroy instanton-antiinstanton pairs. Therefore,
based on these algorithms, it has been attempted to obtain detailed information
on pair distance and $\rho$-distributions from a backward extrapolation of the
cooling/smoothing history~\cite{KOVACS98}.

The second ingredient, which enables us to investigate the 
color correlation between clusters and to examine the coherence inside clusters,
is borrowed from the stochastic vacuum approach~\cite{SIMONOV}.  
In this approach, a gauge invariant field strength correlator (the non-local 
gluon condensate) has been proposed as a model independent characterization 
of the gauge theory vacuum. This correlation function has been studied 
on the lattice~\cite{DIGIACOMO} using the cooling method.
Presently we are calculating~\cite{UNPUBLISHED} the two-point correlator  
\begin{equation}
\label{GGcorr}
D_{\mu\nu\rho\sigma}=\langle~{\mathrm Tr}~\left( 
~G_{\mu\nu}(x_1)~S(x_1,x_2)~G_{\rho\sigma}(x_2)~S(x_2,x_1)~\right)~\rangle 
\end{equation}
where a Schwinger line phase factor 
 $ S(x_1,x_2)=P \exp \left[~i\int_{x_2}^{x_1} dx_{\mu} A_{\mu}(x) \right] $ 
must be inserted. We use CS instead of cooling and the clover construction 
of the field strength components $G^a_{\mu\nu}$. 
In a semiclassical approximation~\cite{MARTEMYANOV}, the leading contribution 
linear in $d(\rho)$ contains just all expressions as given by the classical 
instanton field, for instance - in the regular gauge - $G_{\mu\nu}$ replaced by
\begin{equation} 
\label{FInst}
F^{I,a}_{\mu\nu} = - R^{ab}~\eta^b_{\mu\nu}
		   ~\frac{4~\rho^2}{(|x-z|^2+\rho^2)^2} \, . 
\end{equation} 
Here $z$ and $\rho$ have the meaning explained before, $\eta^a_{\mu\nu}$ is
the 'tHooft symbol and $R^{ab}=\frac12~{\mathrm Tr}~(\tau^aU\tau^bU^+)$ is
the adjoint representation of the global color orientation $U$ of the instanton. 
To obtain the correlator (\ref{GGcorr}), the instanton center $z$ and 
radius $\rho$ have to be averaged over with the density (\ref{density})
after the phase factor $S(x_1,x_2)$ has been evaluated. 

In the present work, once the centers of topological charge clusters 
are identified, the distribution of relative color orientations can be 
obtained by measuring (\ref{GGcorr}) between the centers for pairs of clusters. 
The classical coherence of a cluster can be examined by measuring 
(\ref{GGcorr}) between the center and some other point at a distance up to
a few times $\rho$. While in the model the Schwinger lines
for these cases are radial with respect to the instanton centers 
(and therefore $S(x_1,x_2)=1$), the non-Abelian Schwinger phase
cannot be neglected in the lattice configurations.

In all computations presented here we have used a simplified $SU(2)$
fixed point action on a $12^3\times 4$ lattice~\cite{ILGENFRITZ98},
for which CS as described in~\cite{FEURSTEIN98}
can be carried out in a theoretically consistent manner.
Such actions provide scale invariant ideal instantons~\cite{HASENFRATZ94}
and suppress dislocations~\cite{DEGRAND96}.
Instantons with $\rho < 1.75~a$ (somewhat smaller than the 
blocked level lattice spacing) are instable and cannot be made visible by CS.
Our simulations were carried out in the confinement phase 
($\beta=1.4$ corresponding to a temperature $0.71\,T_c$)   
and slightly above the critical temperature $T_c$ 
($\beta=1.6$, $1.15$ $T_c$).  
To identify instantons in smoothed but true quantum configurations we first 
looked for clusters of topological charge formed by lattice points with a 
charge density of equal sign above a certain threshold 
$q_{\mathrm{thr}}=0.015/a^4$ at $\beta=1.4$ 
and $0.005/a^4$ at $\beta=1.6$. Next, {\it connected} clusters 
are defined by a site-percolation routine. 
The search for local maxima $q_{\mathrm{max}}=\max_x(q(x))$ inside the
clusters identifies the instanton centers $z_i$. The instanton size can be 
obtained from $q_{\mathrm{max}}$
\begin{equation}
\label{rho}
\rho_i=\left(\frac{6}{\pi^2~|q(z_i)|}\right)^{\frac{1}{4}}\, ,
\end{equation} 
as suggested by the classical profile. 
For all clusters we have measured a {\it cluster charge} 
$Q_{\mathrm{cluster}}=\sum_{x \in \mathrm{cluster}} q_L(x)$ 
and a {\it cluster volume}
$V_{\mathrm{cluster}}=\sum_{x \in \mathrm{cluster}} 1$.
Fig. 1 (top)  shows the histogram of instanton radii according to (\ref{rho}).
The left flank represents the probability to find high values of 
$q_{\mathrm{max}}$. The cut-off at large $\rho$ results from the chosen 
threshold value $q_{\mathrm{thr}}$ applied to the cluster search. 
The observed mean value in Fermi of $\bar \rho_{\beta=1.4}=0.51$~fm 
is in good  accordance with \cite{FORCRAND97} where a 
similar $\rho$ definitions was applied.
It is larger than $\bar \rho$ reported in \cite{KOVACS98}. 
In deconfinement we find  $\bar \rho_{\beta=1.6}=0.40$~fm.
Note that in the deconfinement an $O(4)$ asymmetric profile becomes 
important, which might explain the unexpected shape 
of the $\rho$ distribution~\footnote{In present calculations we employ a
characterization of cluster shape in the deconfinement 
that takes care of that.}.
Fig. 1 (bottom) shows how the observed clusters are clustering in the 
$V_{\mathrm{cluster}}$--$Q_{\mathrm{cluster}}$ plane.
The solid line represents the relation between the two cluster
properties predicted according to the classical instanton profile
(\ref{FInst}). The chosen threshold topological density has defined
the horizontal scale, the total (unit) 
charge per cluster determines the vertical one.
The dashed line is the result of a (somewhat deliberately chosen) 
Gaussian profile of a unit topological charge object. This curve
makes clear why 
no cluster charge $Q_{\mathrm{cluster}} > 0.4$ could be localized  
inside an individual cluster. 
The different performance 
of the two  profiles
gives an indication that in the quantum vacuum the instanton
shape might be modified similarly to the fremon model considered in 
Ref~\cite{DIAKONOV84}.

The coordinates of the cluster centers are used to measure histograms
of relative distance for different pairings of cluster charges. 
The observed number of pairs per bin is compared to a random distance histogram.
The random histogram was obtained by a simulation on 30000 configurations on 
the same lattice where the lattice coordinates of I (A) centers were chosen 
randomly all over the lattice with an appropriate density. 
The ratio is shown in Fig. 2 and represents the
two-cluster distance distributions in both phases.  
(Here no distinction was made between time and spacelike separations).
If there were no repulsion or attraction, $C(d)$ would be equal to  
unity. $C(d)>1$ means attractive interaction (pair formation), 
$C(d)<1$ repulsion. The situation for confinement is 
shown in the upper part. For II (AA) pairs no deviation from a random 
distribution is observed for $d > 3~a$. For the IA pairs there is a somewhat 
stronger repulsion up to $d = 3~a$ than for equal sign pairs. 
In the deconfinement (bottom) a peak is observed for 
the IA distance distribution, signaling the formation of IA pairs with a 
center-to-center distance of 2-3 lattice spacings (about $0.28-0.43$ fm). 
For the same-sign case (II or AA) there is a slight repulsion at short 
distances. This IA distance distribution might explain the drop of the 
topological susceptibility $\chi$ at the deconfinement 
phase transition as resulting from a local screening of topological 
charge. In our case  $\chi_{\beta=1.4}= 1.0\cdot 10^{-3}/a^4$  dropped to 
$\chi_{\beta=1.6}= 5.8\cdot 10^{-5}/a^4$ over the transition.  

We have modified the gauge invariant field strength correlator to a 
{\it normalized two-cluster overlap function} for any pair of clusters 
\begin{equation}
\label{twocluster}
{\cal O} = \frac{
		 \langle 
		 {\mathrm Tr}\left(G_{\mu\nu}(x_1)~S(x_1,x_2)
			         ~G_{\mu\nu}(x_2)~S(x_2,x_1)\right)
				 \rangle_{\mathrm{pathes}} 
				 }
                {
		 ({\mathrm Tr}\left(G_{\rho\sigma }(x_1)^2\right))^{\frac12}
                ~({\mathrm Tr}\left(G_{\tau\lambda}(x_2)^2\right))^{\frac12}
				 } \, ,
\end{equation} 
were $x_1$ and $x_2$ denote the positions of the cluster-maxima. The 
Schwinger lines were chosen randomly among the shortest pathes connecting 
$x_1$ and $x_2$ and the average was done incoherently as indicated in 
(\ref{twocluster}). For a non-correlated ensemble of well-separated ideal 
instantons (II) or antiinstantons (AA) one would expect that ${\cal O}$
averages to zero.
In fact, due to the normalization (\ref{twocluster}) 
takes values in the interval 
$(-\frac{1}{3},1)$. This overlap is a measure for
the normalized adjoint trace of the relative gauge orientation between
the clusters,
${\mathrm Tr}_{\mathrm{adj}}~U_{\mathrm{rel}} 
= ({\mathrm Tr}~U_{\mathrm{rel}})^2-1$.
${\cal O}_{II/AA}=-\frac{1}{3}$ corresponding to 
${\mathrm Tr}~U_{\mathrm{rel}}=0$, 
and ${\cal O}=1$ meaning $U_{\mathrm{rel}}=\pm 1$ (color parallelity).
Averaging with the normal Haar measure (no color correlation) would lead 
to $\langle {\cal O} \rangle =0$.
For an IA pair ${\cal O}_{IA}$ is expected to be trivially zero, simply  
due to the opposite
duality
in $\eta^a_{\mu\nu} \bar \eta^b_{\mu\nu} =0$, whatever the relative
orientation is. In order to get access to the relative {\it color} orientation,
in spite of this {\it color blindness} of (\ref{twocluster}), 
we perform a twist in the field strength of one of the clusters in the
IA pair, changing $\vec E^a \longrightarrow -\vec E^a$ before calculating the 
overlap (\ref{twocluster}).  

In order to define $U_{\mathrm{rel}}$, the relative orientation 
between  the clusters, we introduce an $SU(2)$ matrix $U$ into 
(\ref{twocluster}) and try to find, for each pair of clusters, the maximum 
of the function ${\cal O}_{U}$  :
\begin{equation} 
F= \max_{U} {\cal O}_{U} 
 = \max_{U}
        \frac{
	 \langle 
	 {\mathrm Tr}\left(G_{\mu\nu}(x_1)~U~S(x_1,x_2)
		         ~G_{\mu\nu}(x_2)~S(x_2,x_1)~U^+\right)
				 \rangle_{\mathrm{pathes}} 
				 }
                {
		({\mathrm Tr}\left(G_{\rho\sigma }(x_1)^2\right))^{\frac12}
               ~({\mathrm Tr}\left(G_{\tau\lambda}(x_2)^2\right))^{\frac12}
		}  .
\end{equation} 
We find $U$ as the rotational matrix needed to produce maximum overlap, 
{\it i.e.} to undo the relative color rotation $U_{\mathrm{rel}}$.
We found  that ${\cal O}$  {\it measured} for a cluster pair 
can be parametrized as 
${\cal O}=\frac{1}{3}~\left(~4~{\mathrm cos}^2(\alpha)-1\right)$ 
in terms of ${\mathrm cos}(\alpha)=\frac{1}{2}~{\mathrm Tr}~U$ 
{\it as found by maximization} with  very good accuracy, for instance 
$\langle~{\cal O}-\frac{1}{3}~\left(~4~{\mathrm cos}^2(\alpha)-1\right)~
\rangle=-0.02\pm 0.14$  
for $\beta=1.4$.
In about $95\%$ of all pairs of clusters, a rotation could be found turning 
$F$ almost to unity. Only $5.4\%$ of all clusters in the confinement phase 
were so incoherent that they reached  values of $F<0.8$.     
This is indirect support for the picture that the vast majority of the 
identified topological clusters contain all field strength components 
highly coherent in their center.   
We show in Fig. 3 the distributions in 
$\frac{1}{2}~{\mathrm Tr}~U_{\mathrm{rel}}$ for all pairs of clusters with a 
distance $d \leq  4~a$, found in the Monte Carlo sample in the confinement and 
in the deconfinement phase, compared with pairs with $d \leq 5~a$.
The dashed  line shows the distribution according 
to the normal Haar measure (no color correlation) which would lead to vanishing 
average of ${\cal O}$.
The result is surprising in so far as 
in the confinement phase,
opposite charge pairs are nearly uncorrelated while equal charge 
pairs correlated towards color (anti-)parallelity. 
The orientational alignment
becomes stronger in the deconfined phase and is there 
roughly equal for both types
of pairs. We are presently trying to identify the interaction parameters 
of $S_{\mathrm{int}}$ in (\ref{instZ}) by
instanton Monte Carlo simulations.

Finally we will examine the question to what extent the clusters themselves 
can be considered as classical, coherent fields also {\it outside} of their 
centers. It is assumed in the instanton model that, throughout the entire 
instanton, the color orientation of all field strength components follows
the same $U$ in (\ref{FInst}). Corroborating what we have discussed in
relation to
Fig. 1 (bottom), 
we find here that the semiclassical picture has to be revised since 
the two-point field strength correlator inside a cluster has a profile which 
differs from the classical one. In order to define a correlation length within 
a topological cluster, we have modified the field strength correlator to a 
{\it normalized cluster profile} 
\begin{equation}
\label{onecluster}
{\cal P} = \frac{
		 {\mathrm Tr}\left(G_{\mu\nu}(x_1)~S(x_1,x_2)
			         ~G_{\mu\nu}(x_2)~S(x_2,x_1)\right)
				 }
                {
		 {\mathrm Tr}\left(G_{\rho\sigma}(x_1)^2\right)
				 }
\end{equation} 
pinning $x_1$ to the center of a cluster and considering it as a function
of distance $d=|x_2-x_1|$. 
For these calculations we restricted ourselves to on-axis displacements
$x_2-x_1$ such that no path averaging was necessary.  No twist was applied 
to the field strength at $x_2$ such that fields of 
opposite duality could not contribute. The resulting profile function 
is shown in Fig. 4 in the confinement (left) and 
deconfinement phase (right).
Before final conclusions can be drawn the statistics of
analyzed clusters has to be accumulated. 
The dashed  line drawn in both figures represents a fit by 
a Gaussian profile function ${\cal P}(x)=\exp(-x^2/2\tilde{B})$.
The squares show an average over the field strength profile of 
individual clusters described by
$q_{\mathrm{cluster}}(x)=\exp(-x^2/B)/(\pi^2~B^2)$,
averaged over the empirical $B$ distribution obtained from $q_{\mathrm{max}}$. 
The solid line is the best fit to the average using the classical profile. 
This comparison substantiates our tentative model 
introduced to explain the $Q_{\mathrm{cluster}}$--$V_{\mathrm{cluster}}$ 
correlation in the confinement phase. Let us define a {\it r.m.s.} size
(for weighting with respect to the topological density) of the instanton 
$\rho_{\mathrm{rms}}=\sqrt{2~B}$. From the $B$ distribution we obtain
the average $\overline{B}=2.71~a^2$ at $\beta=1.4$. 
If the string tension is taken to be $\sqrt{\sigma}= 440$ MeV 
we find at $\beta=1.4$ a lattice spacing $a=0.23$~fm which corresponds to
an $\rho_{\mathrm{rms}}=0.536$ fm. 
At $\beta=1.6$ we find 
an average $\overline{B}=2.87~a^2$. 
An extrapolation using the 2-loop formula 
gives $a=0.142$ fm at $\beta=1.6$ which results in 
$\rho_{\mathrm{rms}}=0.340$~fm.
These numbers are very  similar to the average radius defined via
$\rho$ according to (\ref{rho}) and its distribution, although the
modification of the profile at large distances is of principal importance
for the confinement phase.
We have to stress the possibility of asymmetry effects changing the instantons 
in the deconfined phase. This was not yet taken into account in this 
feasibility study. We are currently studying shape and correlations
by treating time and space directions separately.  

In conclusion we emphasize that the use  of constrained 
smoothing in combination with the measurement of gauge 
invariant two-point field strength
correlators opens the possibility to obtain 
essential input information for semi-classically motivated models of QCD 
which was not available before.

We thank D. Diakonov and M. M\"uller-Preussker for initializing 
discussions and continued interest. 
\newpage

\newpage

\begin{figure}
\caption{   
Top:  
Instanton size distribution according to (\ref{rho}) in lattice spacings. 
The threshold values were $q_{\mathrm thr}=0.015$ in the confinement (left) 
and $0.005$ in the deconfinement phase (right) such that
instantons with radius $\rho > 2.523~a$ or $3.321~a$
are not visible. 
Bottom: 
average and variance of a scatterplot of $Q_{\mathrm{cluster}}$ vs. 
$V_{\mathrm{cluster}}$. The agreement with the curve predicted by the
classical profile (solid line) is lost for $V_{\mathrm{cluster}}>8$. 
A Gaussian  instanton profile (dashed line) fits the data better
in the confinement phase. 
Note that only $6.4\%$ of the total number of clusters 
analyzed had a cluster volume $V_{\mathrm{cluster}}> 8$. 
}
\end{figure}

\begin{figure}
\caption{Distance distribution $C(d)$ for equal charge (II and AA) and 
opposite charge (AI) pairs 
in the confinement (top) and deconfinement phase (bottom). 
The peak in the deconfinement 
phase for AI pairs indicates pair formation. 
Results obtained from 8930 clusters (1000 independent 
configurations) at $\beta=1.4$ and 8862 clusters (2000 configurations) 
at  $\beta=1.6$. }  
\end{figure}

\begin{figure}
\caption{Distribution in ${\mathrm Tr}~U_{\mathrm{rel}}$ for II(AA) and AI pairs 
in the confinement (top) and deconfinement phase (bottom)
averaged over all relative displacements $d \leq 4~a$ (fat line) and 
$d \leq 5~a$ (thin line). 
The random distribution (dashed line) is shown for comparison.
 }  
\end{figure}

\begin{figure}
\caption{
Cluster profile ${\cal P}$ (circles) obtained according to (\ref{onecluster})  
from 100 independent configurations 
containing $O(1000)$ clusters  
at $\beta=1.4$ (left) and
containing $O(300)$ clusters at $\beta=1.6$ (right). 
Solid lines are fits to the classical profile with free parameter $\rho$, 
dashed curves correspond to the best fit to a Gaussian profile with free 
parameter $\tilde B$.  
Squares represent the averages over Gaussian profiles
according to a $B$ distribution obtained from $q_{\mathrm max}$ 
for each individual cluster. 
}  
\end{figure}

\newpage
\pagestyle{empty}

\begin{figure}
\begin{tabular}{cc}
{\LARGE $\beta=1.4$}  & {\LARGE $\beta=1.6$}\\
\epsfxsize=8.5cm\epsffile{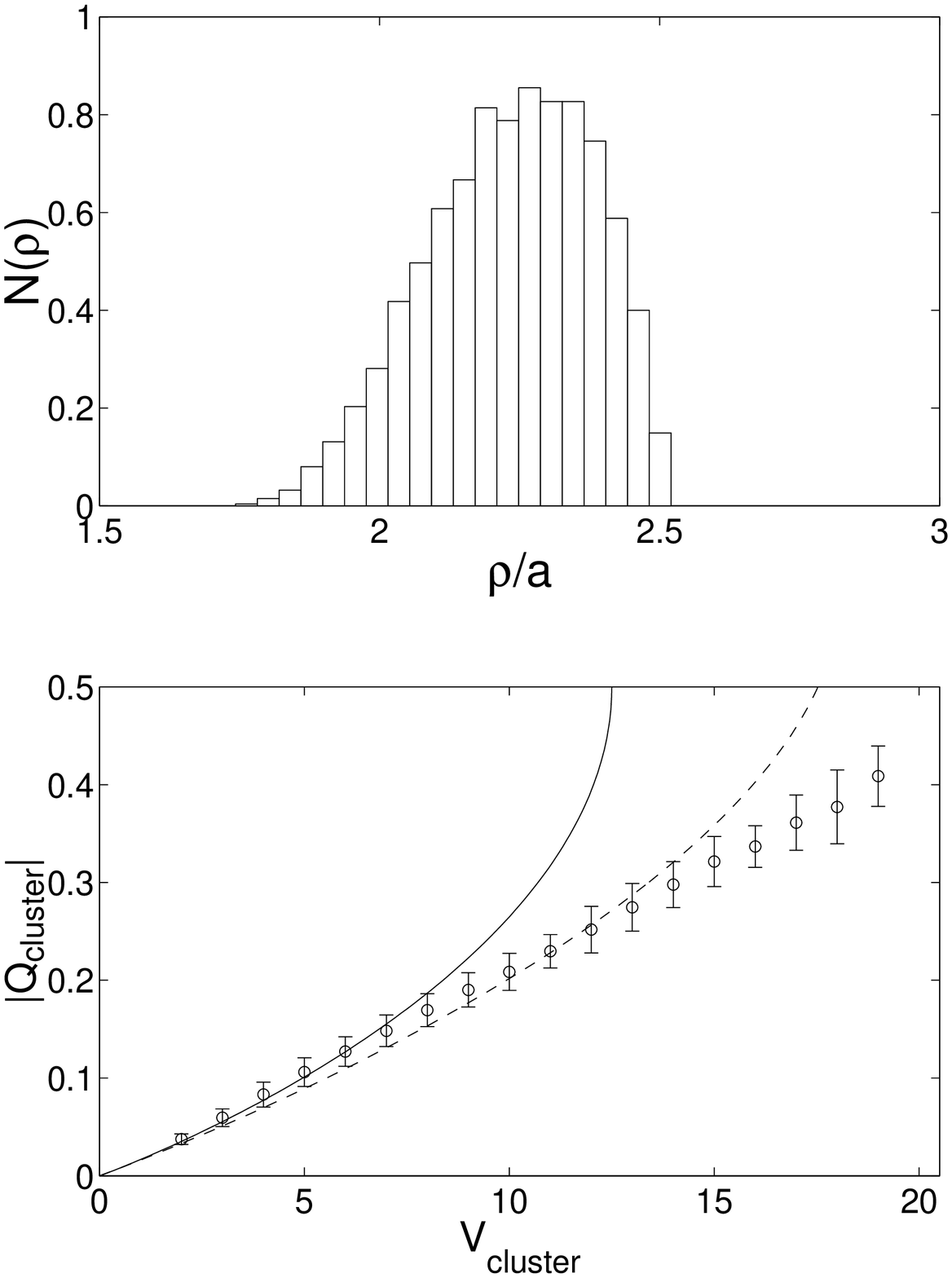} & 
\epsfxsize=8.5cm\epsffile{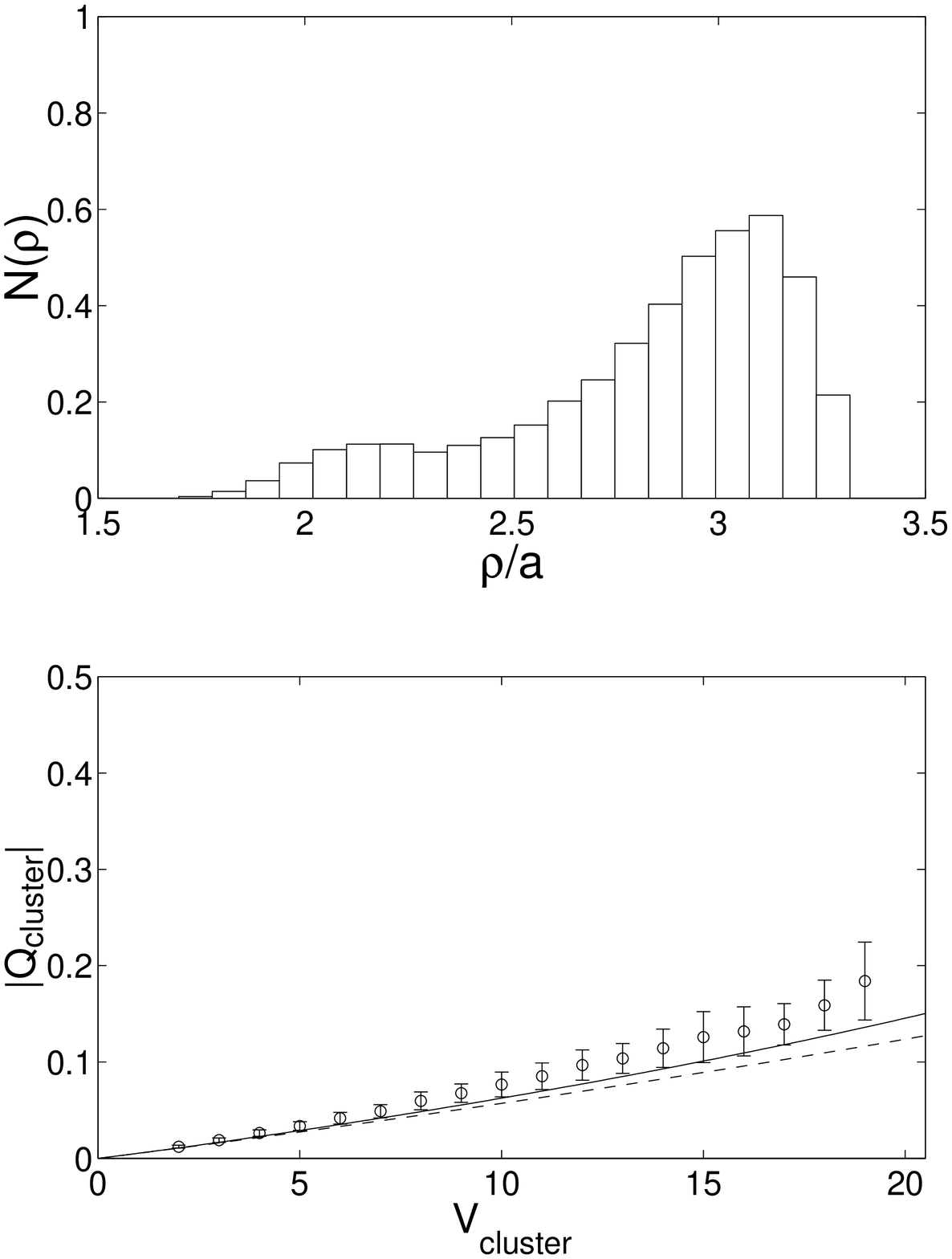} 
\end{tabular}
\end{figure}

\vfill
\begin{center}
FIG. 1
\end{center}

\newpage

\vspace{3cm}

\begin{figure}
\epsfxsize=17.3cm\epsffile{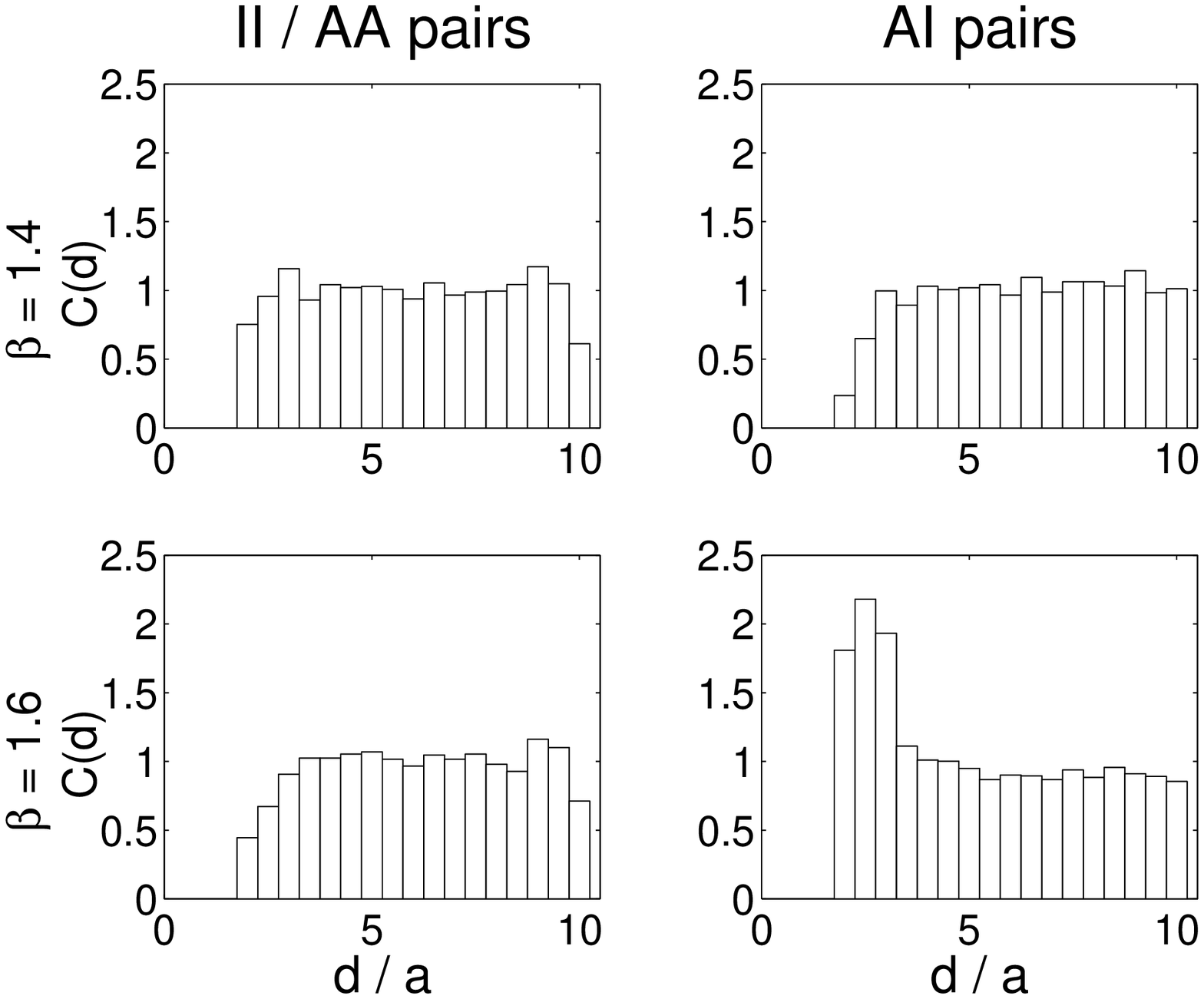}
\end{figure}

\vfill
\begin{center}
FIG. 2
\end{center} 

\newpage

\begin{figure}
\epsfxsize=17.3cm\epsffile{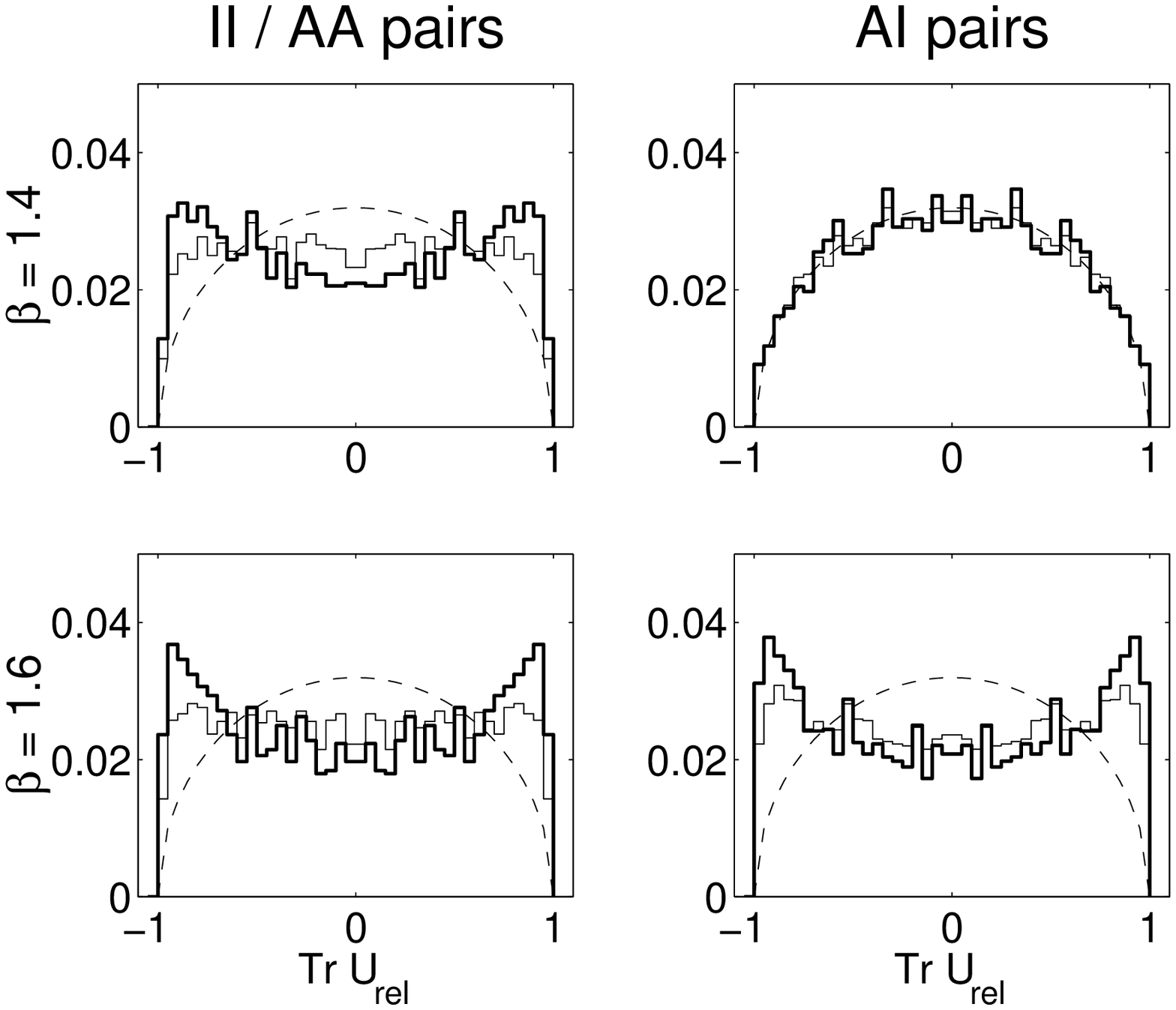}
\end{figure}

\vfill
\begin{center}
FIG. 3
\end{center} 

\newpage

\begin{figure}
\begin{tabular}{cc}
\hspace{5mm} {\LARGE $\beta=1.4$ }&\hspace{5mm}  {\LARGE $\beta=1.6$ }\\
\epsfxsize=8.0cm\epsffile{profil_fit_sm_140.eps2} & 
\epsfxsize=8.0cm\epsffile{profil_fit_sm_160.eps2} \\
\end{tabular}
\end{figure}

\vfill
\begin{center}
FIG. 4
\end{center} 

\newpage 


\end{document}